\documentclass[aps,prl,twocolumn,showpacs,floatfix,superscriptaddress]{revtex4}
\usepackage{graphicx}
\usepackage{amssymb}
\usepackage{textcomp}
\usepackage{bm}

\begin{document}
\setcounter{page}{1}
\title[]{Anomalous Corrections to Hall Resistivity of Spin-Polarized Two-Dimensional Holes
in a GaAs/AlGaAs Heterostructure}

\author{Hwayong Noh}
\email{hnoh@sejong.ac.kr}
\affiliation{Department of Physics and Institute of Fundamental Physics, 
Sejong University, Seoul 143-747, Korea}

\author{S. Lee}
\affiliation{Department of Physics and Institute of Fundamental Physics, 
Sejong University, Seoul 143-747, Korea}

\author{S. H. Chun}
\affiliation{Department of Physics and Institute of Fundamental Physics, 
Sejong University, Seoul 143-747, Korea}

\author{H. C. Kim}
\affiliation{National Fusion Research Institute, Daejon 305-333, Korea}

\author{L. N. Pfeiffer}
\affiliation{Bell Labs, Alcatel-Lucent, Murray Hill, New Jersey 07974, USA}

\author{K. W. West}
\affiliation{Bell Labs, Alcatel-Lucent, Murray Hill, New Jersey 07974, USA}

\date{\today}

\begin{abstract}
Hall effect of two-dimensional holes that are spin-polarized by a strong parallel
magnetic field was explored with a small tilt angle.
The Hall resistivity increases nonlinearly with the magnetic field and
exhibits negative corrections. 
The anomalous negative corrections increase with the perpendicular magnetization of 
the two-dimensional hole system. We attribute this to the anomalous Hall effect of 
spin-polarized carriers in a nonmagnetic system. The anomalous corrections to the
conductivity exhibit non-monotonic dependence on the magnetic field.
\end{abstract}


\maketitle

Hall effect in a ferromagnetic system exhibits anomalous contributions resulting 
from the spin polarization of the carriers and the spin-orbit coupling. 
First observed in ferromagnetic metals, the so-called 
anomalous Hall effect(AHE)\cite{ahe} has gained new attention with the
development of diluted magnetic semiconductors(DMS) and the observation of the 
ferromagnetic transition and AHE in them\cite{dms}. 
It was also observed in a paramagnetic DMS system\cite{cumings}, 
where the added magnetic impurities enhance the g-factor of the electrons 
to make them spin-polarized with a small magnetic field. 
Extending the idea to a nonmagnetic system without intentional magnetic impurities, 
the AHE should be observable once the carriers are made spin-polarized by some means, 
e.g. strong enough magnetic fields. Doing this, however, is not easy for most of
the systems because the magnetic field required is too large, while
for low density two-dimensional(2D) carrier systems, it is possible to spin-polarize
the system with a moderately high magnetic field. Probing the AHE in a nonmagnetic 2D 
carrier system has some other importance as well.
The 2D carrier systems in GaAs/AlGaAs heterostructures, one of the most widely studied
systems, usually have very high mobilities, and therefore the AHE in such systems
could be dominated by an intrinsic origin.

In this paper, we report the Hall measurements on a low density 2D hole 
system in a GaAs/AlGaAs heterostructure that was spin-polarized by parallel 
magnetic fields. By tilting the sample slightly from the position where the magnetic 
field is parallel to the 2D plane, we generated a tiny perpendicular component of 
the field and measured the resulting Hall voltages. The measured Hall resistivity 
does not increase linearly with the magnetic field and the Hall slope exhibits negative
corrections going through a local minimum as the 2D holes are spin polarized. 
Analyses on these negative corrections reveal behaviors that are correlated with the degree
of magnetization perpendicular to the 2D plane and that are attributable to the anomalous
Hall effect. For a fixed tilt angle, the negative corrections
increase with the increasing magnetic field and saturate above the full polarization
field. They also increase with the tilt angle for a fixed magnetic field.
Extracting the corrections in conductivity yields a more surprising non-monotonic
dependence on the magnetic field, a possible evidence for the intrinsic effect.

The sample used is a Hall-bar shaped 2D hole system in an undoped (100) GaAs/AlGaAs 
heterostructure\cite{kane}, where the 2D holes are induced by a metallic gate 
on top of the structure.
The mobility of the holes measured at a temperature($T$) of 0.3 K is 
$2.9\times10^{5}$ cm$^{2}$/Vs for the hole density $p=2.8\times10^{10}$ cm$^{-2}$.
Measurements were done in a He-3 refrigerator with a base temperature of 0.3 K.
The sample was mounted in a rotation stage so that the tilt angle between
the 2D plane and the magnetic field($B$) could be adjusted {\it in-situ}.
The longitudinal($\rho_{xx}$) and the Hall resistivity($\rho_{xy}$) under the
tilted $B$ were 
measured by the standard lock-in technique with an excitation current of 10 nA.
To remove the effects of misaligned contacts, $\rho_{xy}$ was obtained from two
measurements with opposite directions of the $B$.

The position where the $B$ is parallel to the 2D plane was accurately determined 
by making the Hall voltage as small as possible at the highest $B$ of 7 T.
When the sample was tilted from the parallel position, the Hall voltages could
not be used to determine the tilt angle due to the existing corrections in the Hall 
voltages. Shubnikov-de Haas(SdH) oscillations could not be used either
since we limited the tilt angle below $1.05^{\circ}$ to avoid such oscillations.
Therefore, we used a monitor sample mounted intentionally tilted relative 
to the 2D hole sample for the angle measurement.
The monitor sample was a 2D electron system in another GaAs/AlGaAs 
heterostructure with the electron density of $n=3.7\times10^{11}$ cm$^{-2}$.
With the 2D hole sample in the parallel position, the SdH oscillations from
the monitor sample yielded the angle between the two, which is $5.58^{\circ}$. 
When the 2D hole sample was tilted from the parallel position, the tilt angle
of the monitor sample extracted from the SdH oscillations minus the angle between
the two samples gave the tilt angle of the 2D hole sample, $\theta$.

\begin{figure}
\begin{center}
\includegraphics[width=2.9in]{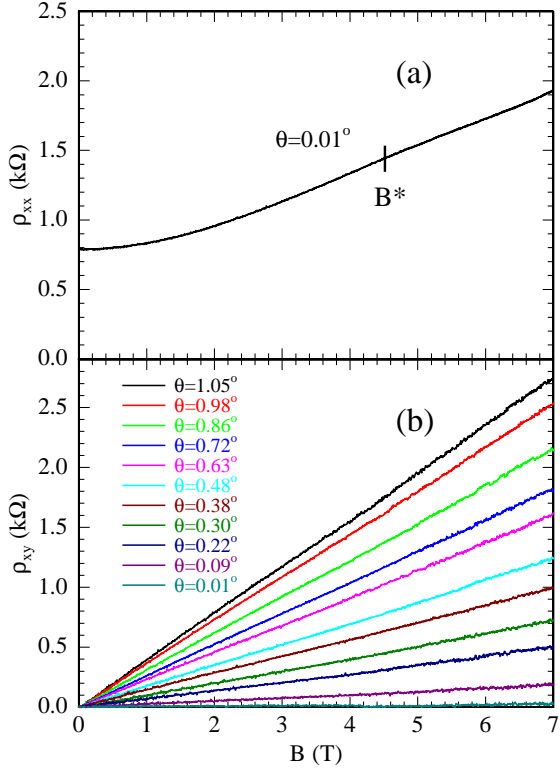}
\end{center}
\caption{(Color online) (a) $\rho_{xx}$ vs $B$ for $\theta=0.01^{\circ}$, where $B$ 
is almost parallel to the 2D hole layer. There is a slight bump around 4.5 T, which is
indicated by $B^{*}$. (b) $\rho_{xy}$ vs $B$ for different $\theta$.
The 2D hole density is $p=2.8\times10^{10}$ cm$^{-2}$} \label{1}
\end{figure}

Figure~\ref{1} (a) shows $\rho_{xx}$ under the parallel $B$ 
($\theta=0.01^{\circ}$), and Fig.~\ref{1} (b) shows $\rho_{xy}$ measured 
as a function of $B$ for different $\theta$.
For $\theta=0.01^{\circ}$, $\rho_{xx}$ increases monotonically with $B$ and there is a 
slight change in the $B$-dependence of $\rho_{xx}$ characterized by a bump around
$B^{*}=4.5$ T. This feature has been observed in many other low density 2D systems, 
and is associated with the full spin polarization (or spin subband depopulation)\cite{mr}.
In other words, only one spin subband is occupied above this field. 
$\rho_{xy}$ for $\theta=0.01^{\circ}$, on the other hand, 
does not increase much up to $B=7$ T, becoming about 30 $\Omega$.
When the sample is tilted from the parallel position, a perpendicular 
component($B_{\perp}$) of $B$ is generated, and accordingly Hall voltages
develop. Since $B_{\perp}=Bsin\theta$, $\rho_{xy}$ increases faster for larger $\theta$.
We limited $\theta$ below $1.05^{\circ}$ so that $B_{\perp}$ is below 0.13 T
where the SdH oscillations do not develop.

\begin{figure}
\begin{center}
\includegraphics[width=2.9in]{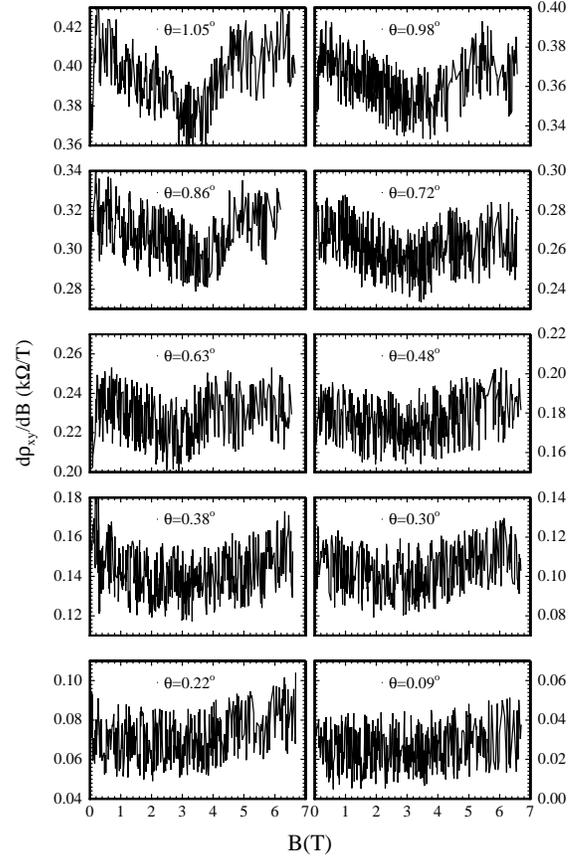}
\end{center}
\caption{The Hall slope, $d\rho_{xy}/dB$, is plotted as a function of $B$ for different 
$\theta$.} \label{2}
\end{figure}

If the Hall voltages that develop are solely from the ordinary Hall effect, $\rho_{xy}$
should increase linearly with $B_{\perp}$, hence with $B$ for a fixed $\theta$.
In Fig.~\ref{1} (b), it appears that way at first sight, but a careful examination
reveals a deviation from the simple linear increase with $B$.
In Fig.~\ref{2}, we show the Hall slope $d\rho_{xy}/dB$ for the same data. The Hall 
slope decrease with $B$, goes through a local minimum, and then increases again at 
higher $B$. For $B$ higher than 4.5 to 5 T, the Hall slope appears to saturate, which
is more evident for larger $\theta$. This field is very close to the spin depopulation 
field $B^{*}$ indicated in Fig.~\ref{1} (a).
 
To figure out the origin of this deviation, we first consider the Hall slope in
the two band model. In our sample, the light and the heavy hole bands are split due to 
the confinement potential of the heterostructure, and only the heavy hole band is
occupied since the hole density is in the low $10^{10}$ cm$^{-2}$ range. The heavy 
hole band itself consists of two spin subbands with spin component $\pm3/2$. 
By the application of in-plane magnetic field, these subbands are split due to the 
Zeeman effect. While the Hall slope for a single band is inversely proportional to 
the carrier density, it is rather complicated for two bands.
Assuming no inter-subband scattering, the Hall slope in this case can be 
written as $(p_{1}\mu_{1}^{2} + p_{2}\mu_{2}^{2})/e(p_{1}\mu_{1}+p_{2}\mu_{2})^{2}$, 
where $p_{1}$ and $p_{2}$ are hole densities in each spin subband, and $\mu_{1}$ and 
$\mu_{2}$ are mobilities. 
For a simple estimation, we can assume that $p_{1}=(p/2)(1+B/B^{*})$ and 
$p_{2}=(p/2)(1-B/B^{*})$, where $p$ is the total density. For the hole density of 
our sample, the mobility decreases with decreasing density by a power 
law, and we can assume that $\mu_{1}\sim p_{1}^{n}$ and $\mu_{2}\sim p_{2}^{n}$. 
Then, the Hall slope exhibits a local maximum for $0<B<B^{*}$. A consideration of 
the inter-subband scattering does not change this behavior even though 
it could suppress the degree of the variation in the Hall slope\cite{vitkalov}.
Therefore, the two band model cannot explain our data that show a local minimum 
in the Hall slope.

The next thing we can consider is the interaction effects.
It has been known that interaction effects give a logarithmic correction to
the longitudinal and the Hall resistivity in the diffusive regime, $k_{B}T\tau/\hbar<1$,
where $\tau$ is the transport scattering time\cite{altshuler}. 
A more recent theory by Zala {\it et al.}\cite{zala}
extended the scope to the ballistic regime, $k_{B}T\tau/\hbar>1$, and showed that
the corrections to the Hall resistivity go as $1/T$.
Since $k_{B}T\tau/\hbar>1$ in our sample, we can estimate the interaction corrections 
in the ballistic regime. There are two contributions in the corrections, 
the singlet($\delta\rho_{xy}^{\rho}$) and the triplet($\delta\rho_{xy}^{\sigma}$)
channel corrections. At $B=0$, both corrections are present, while
at high enough $B$ where spins are polarized, only the singlet corrections remain.
If the deviation in $\rho_{xy}$ observed in our experiment is related to the interaction
effects, the amount of the deviation is presumably the same as
the triplet corrections that disappear at high $B$. 
To calculate the triplet corrections, we first found the Fermi liquid interaction
parameter, $F_{0}^{\sigma}$, by fitting the $\rho_{xx}$ data in Fig.~\ref{1} (a)
below 0.7 T with the formula given by Zala {\it et al.}. 
This gives $F_{0}^{\sigma}=-0.2$, and in turn we get
$\delta\rho_{xy}^{\sigma}/\rho_{xy}=-0.0005$. This value is not only too small
in magnitude but also has an opposite sign. The deviation we observed at high $B$
is about 6$\sim$20 \%, hundreds times larger than the triplet corrections.
In addition, if the negative triplet corrections that are present at low $B$
disappear at high $B$, the Hall slope should go through a local maximum.
Therefore, the interaction corrections cannot explain the data either.

\begin{figure}
\begin{center}
\includegraphics[width=2.8in]{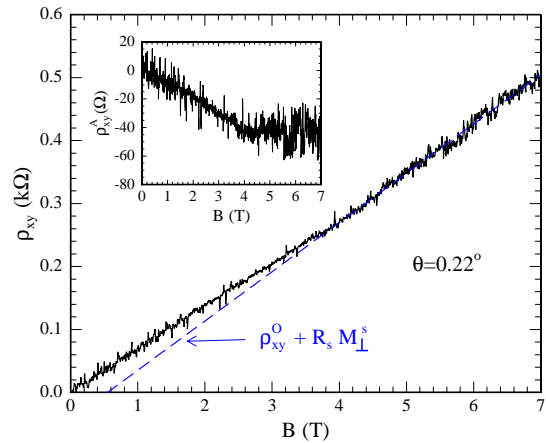}
\end{center}
\caption{(Color online) $\rho_{xy}$ vs $B$ for $\theta=0.22^{\circ}$. Blue dashed line
is a linear fit to the data for $B>4.5$ T. Inset: $\rho_{xy}^{A}=\rho_{xy}-\rho_{xy}^{O}$
vs $B$.} \label{3}
\end{figure}

This leads us to see a possibility that the deviation might come from the
anomalous Hall effect. Although the 2D hole system in our experiment is paramagnetic,
it would behave similarly to a ferromagnet when the spins are fully polarized. 
In fact, the work by Cumings {\it et al.}\cite{cumings} reported the AHE
in a paramagnetic 2D electron system, where the 2D electrons are spin polarized
under a small perpendicular $B$. Their sample, however, contained
a magnetic alloy of Mn, while the 2D hole sample used in our experiment
does not contain any intentional magnetic impurities. If the deviation 
is indeed due to the AHE, we can reexamine the Hall data of Fig.~\ref{1} (b)
in the following respect. 
The Hall resistivity $\rho_{xy}$ has an ordinary and an anomalous contribution,
and can be represented by $\rho_{xy}=\rho_{xy}^{O}+R_{s}M_{\perp}$, 
where $\rho_{xy}^{O}$ is the ordinary Hall resistivity, $R_{S}$ the anomalous Hall 
coefficient, and $M_{\perp}$ the perpendicular magnetization of the 2D hole system.
While $\rho_{xy}^{O}$ increases linearly with $B$, $M_{\perp}$ will increase with $B$ 
until $B$ reaches the depopulation field and will saturate to a value $M_{\perp}^{s}$. 
Therefore, $\rho_{xy}$ above 4.5 T can be expressed by $\rho_{xy}^{O}+R_{s}M_{\perp}^{s}$.
We show an example for $\theta=0.22^{\circ}$ in Fig.~\ref{3}.
$\rho_{xy}$ is well fitted by a straight line for $B>4.5$ T (blue dashed line), and
there is a clear deviation appearing at low $B$ implying a negative value of $R_{s}$.
The difference between $\rho_{xy}$ and $\rho_{xy}^{O}$, which can be attributed to
the anomalous Hall resistivity $\rho_{xy}^{A}=R_{s}M_{\perp}$, is plotted in the inset. 
It is negative, and its magnitude increases with $B$ before saturating to a value that 
corresponds to $R_{S}M_{\perp}^{s}$, suggesting that the deviation is correlated with 
$M_{\perp}$. The fact that the deviation in the Hall slope is larger for larger $\theta$ 
in Fig.~\ref{2} provides an additional support for the correlation with $M_{\perp}$
since $M_{\perp}$ will increase with $\theta$.

\begin{figure}
\begin{center}
\includegraphics[width=3in]{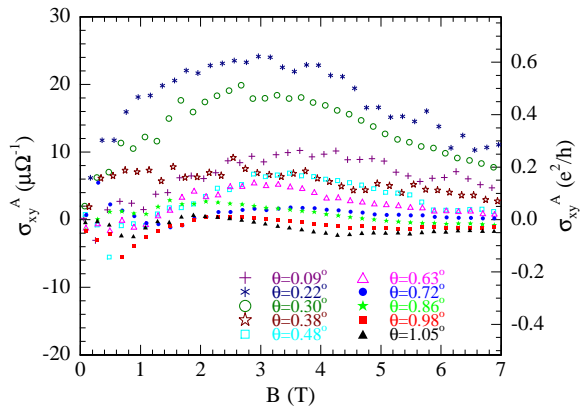}
\end{center}
\caption{(Color online) $\sigma_{xy}^{A}=\sigma_{xy}-\sigma_{xy}^{O}$ vs $B$ for different
$\theta$. $\sigma_{xy}$ is obtained from $\rho_{xx}$ and $\rho_{xy}$, and
$\sigma_{xy}^{O}$ is from $\rho_{xx}$ and $\rho_{xy}^{O}$.} \label{4}
\end{figure}

There had been many extensive theoretical studies on the AHE of 
2D electron systems with the Rashba spin-orbit coupling. Most recently, the work by 
Nunner {\it et al.}\cite{nunner} and Kato {\it et al.}\cite{kato} provided very 
thorough calculations of the anomalous Hall conductivity.
Although we cannot make a direct comparison with those theoretical calculations
which consider an electron system and somewhat different models incorporating an 
exchange field, we still believe that it is stimulating to contrast our data with those 
theories. For this, we first calculated 
Hall conductivity($\sigma_{xy}$) from the measured $\rho_{xx}$ and $\rho_{xy}$. 
Then, to extract the anomalous Hall conductivity($\sigma_{xy}^{A}$), 
we subtracted the ordinary Hall
conductivity($\sigma_{xy}^{O}$), which was calculated from $\rho_{xx}$
and $\rho_{xy}^{O}$. The results are shown in Fig.~\ref{4}.

$\sigma_{xy}^{A}$ as a function of $B$ for different $\theta$ exhibits somewhat
complicated behavior. However, one important feature is that $\sigma_{xy}^{A}$
has a non-monotonic dependence on $B$. It increases at low $B$, goes through a
maximum, and then decreases at higher $B$. This non-monotonicity is surprisingly
similar to that found in the numerical calculations by Kato {\it et al.}\cite{kato} and 
other previous calculations on the intrinsic anomalous Hall conductivity\cite{culcer}. 
In those calculations, the intrinsic anomalous Hall conductivity of 2D electrons with 
Rashba spin-orbit coupling shows a non-monotonic dependence on $\Delta_{ex}/E_{F}$,
where $\Delta_{ex}$ is the exchange splitting and $E_{F}$ is the Fermi energy. 
It was pointed out\cite{kato,nunner}, however, that the anomalous Hall conductivity 
of 2D electrons vanishes when $\Delta_{ex}/E_{F}<1$ if the scattering time is 
spin-independent.
In our sample, $\sigma_{xy}^{A}$ is nonzero for $B<4.5$ T, where the Zeemann
splitting, which is basically $\Delta_{ex}$, is less than $E_{F}$. 
Although some origins for the spin-dependent scattering time would be possible,
the different nature of the Rashba spin-orbit coupling in the 2D hole systems 
could be also playing an important role.
Since the Rashba spin-orbit splitting is third order in $k$\cite{winkler}
for the 2D hole systems, the intrinsic anomalous Hall 
conductivity is not necessarily canceled by the disorder effect even though
the scattering time is spin-independent. 
Moreover, the Rashba spin-orbit splitting $\alpha k_{F}^{3}$ 
is larger than $\hbar/\tau$ in our sample\cite{winkler2}, and therefore 
the system is in the clean limit, a favorable condition to observe the intrinsic
effect. Finally, Borunda {\it et al.}\cite{borunda} predicted that the skew 
scattering, a principal extrinsic contribution for systems with low disorder, 
is absent for 2D hole systems. Therefore, the non-monotonic behavior of 
$\sigma_{xy}^{A}$ observed in our experiment could be a strong evidence for
the intrinsic AHE.

In Fig.~\ref{4},
not only the non-monotonic behavior of $\sigma_{xy}^{A}$ but also the magnitude
is similar to that in the numerical calculations, being several tenths of $e^{2}/h$.
However, what makes the magnitude and the peak position different for different 
$\theta$ cannot be understood at this time. 
In the calculations by Kato {\it et al.}\cite{kato}, different values of 
spin-orbit coupling gave such differences. Rashba spin-orbit coupling does not
change when the sample is tilted. Instead, the perpendicular magnetization
increases with the tilt angle. If the different values of perpendicular magnetization
might be considered as an effective change of $\Delta_{ex}$, 
it could affect the anomalous Hall conductivity. 
We can also conjecture that the perpendicular spin polarization of unintentional magnetic 
impurities, if exist, changes with $\theta$ resulting in the change of the anomalous
Hall conductivity\cite{nunner2}. A more detailed calculation appropriate to our 
experimental situation would be needed. 

In summary, the Hall resistivity of 2D holes in a GaAs/AlGaAs heterostructure 
under a slightly tilted-from-parallel magnetic field shows negative corrections.
These anomalous corrections increase with increasing perpendicular magnetization
of the 2D hole system. In terms of conductivity,
the anomalous corrections, being several tenths of $e^{2}/h$ in magnitude,
show non-monotonic dependence on the magnetic field, a behavior expected for
the intrinsic AHE.
 
We would like to thank M. P. Lilly and D. C. Tsui for their initial support for
making the sample.
This work was supported by the Korea Research Foundation Grant funded by the Korean 
Government (MOEHRD, Basic Research Promotion Fund)(KRF-2008-314-C00139).

\end{document}